%% file: arxiv.tex
\documentclass[aps,prl,twocolumn,showpacs,superscriptaddress,groupedaddress,floatfix]{revtex4}  
\usepackage{graphicx}  
\usepackage{dcolumn}   
\usepackage{bm}        
\usepackage{amssymb}   
\usepackage{multirow}
\usepackage{verbatim}
\usepackage{array}
\usepackage{subfigure}

\def\met{\ensuremath{E\kern-0.57em/_{T}}}
\def\pet{\ensuremath{p\kern-0.57em/_{T}}}

\begin{document}
\hspace{5.2in} \mbox{Fermilab-Pub-11/267-E}
\title{Bounds on an anomalous dijet resonance in $W+$~jets production in \\
\boldmath{$p\bar{p}$} collisions at $\sqrt{s} =1.96$~TeV}

\input author_list.tex

\date{June 9, 2011}

\begin{abstract}
We present a study of the dijet invariant mass spectrum in events with
two jets produced in association with a $W$ boson in data
corresponding to an integrated luminosity of 4.3~fb$^{-1}$ collected
with the D0 detector at $\sqrt{s} =1.96$~TeV.  We find no evidence for 
anomalous resonant dijet production and derive upper limits 
on the production cross section of an anomalous dijet resonance recently 
reported by the CDF Collaboration, investigating the range of dijet 
invariant mass from 110 to 170~GeV/$c^2$.  The probability of the D0 
data being consistent with the presence of a dijet resonance with 4~pb 
production cross section at 145~GeV/$c^2$ is $8\times 10^{-6}$.
\end{abstract}

\maketitle

The CDF Collaboration at the Fermilab Tevatron $p\bar p$ collider
recently reported a study of the dijet invariant mass ($M_{jj}$)
spectrum in associated production with $W\to\ell\nu$ ($\ell=e$ or
$\mu$) at $\sqrt{s}=1.96$~TeV with an integrated luminosity of 4.3
fb$^{-1}$~\cite{bib:CDFbump}.  In that paper they present evidence 
for an excess of events corresponding to 3.2 
standard deviations (s.d.)~above the background expectation, centered at
$M_{jj}=144$~$\pm$~5~GeV/$c^2$~\cite{bib:CDFbump}.  The CDF authors 
model this excess using a Gaussian peak with a width corresponding 
to an expected experimental $M_{jj}$ resolution for the CDF
detector~\cite{bib:CDFdet} of 14.3~GeV/$c^2$ and further estimate the
acceptance and selection efficiencies by simulating associated 
{\sl W}~+~Higgs boson ($H$) production in the decay mode $H \to
b\bar{b}$ and with a mass $M_H=$~150~GeV/$c^2$.  Assuming the excess is
caused by a particle $X$ with $\mathcal{B}(X\to jj)=1$, the CDF Collaboration 
reports an estimated production cross section of 
$\sigma(p\bar{p}\to WX)\approx 4$~pb.

Using 5.3~fb$^{-1}$ of integrated luminosity, the D0 Collaboration has
previously set limits on resonant $b\bar b$ production in association
with a $W$ boson in dedicated searches for standard model (SM) Higgs
bosons in the $WH \to \ell \nu b\bar{b}$ channel~\cite{bib:whlvbbD0}.  The 
D0 Collaboration reported upper limits on 
$\sigma(p\bar{p}\to WH)\times \mathcal{B}(H\to b\bar b)$ ranging from
approximately 0.62~pb for $M_{H} =100$~GeV/$c^2$ to 0.33~pb for
$M_{H}=$~150~GeV/$c^2$.  The CDF Collaboration has
performed a similar search using 2.7~fb$^{-1}$ of integrated luminosity 
and reported no excess of events~\cite{bib:whlvbbCDF}.  Furthermore, 
the D0 Collaboration has not observed a significant excess of associated 
$W$ boson and dijet production in analyses of either 
$WW/WZ\to\ell\nu jj$~\cite{bib:D0prl} or 
$H\to WW\to\ell\nu jj$~\cite{bib:hwwlvjjD0} using 1.1~fb$^{-1}$ and
5.4~fb$^{-1}$ of integrated luminosity, respectively.

In this Letter we report a study of associated $W ( \to \ell\nu)$ and
dijet production using data corresponding to 4.3~fb$^{-1}$ of
integrated luminosity collected with the D0 detector~\cite{bib:detector} 
at $\sqrt{s} =1.96$~TeV at the Fermilab Tevatron $p\bar{p}$ Collider.  
The CDF study of this production process uses the same integrated luminosity.   
We investigate the dijet invariant mass range from 110 to 170~GeV/$c^2$ 
for evidence of anomalous dijet production.

To select $W( \to \ell\nu) + {jj}$ candidate events, we impose similar
selection criteria to those used in the CDF analysis: a single
reconstructed lepton (electron or muon) with transverse momentum $p_T
>20$~GeV/$c$ and pseudorapidity~\cite{bib:def} $|\eta| <1.0$; 
missing transverse energy $\met >25$~GeV; two jets reconstructed 
using a jet cone algorithm~\cite{bib:JetCone} with a cone of radius 
$\Delta\mathcal{R}=0.5$ that satisfy $p_T>30$~GeV/$c$ and $|\eta| <2.5$, 
while vetoing events with additional jets with $p_T >30$~GeV/$c$.  
The separation between the two jets must be 
$|\Delta\eta(\text{jet}_{1},\text{jet}_{2})| <2.5$, and the azimuthal 
separation between the most energetic jet and the direction of the $\met$ 
must satisfy $\Delta\phi(\text{jet},\met) >0.4$.  The transverse momentum of the
dijet system is required to be $p_T(jj) >40$~GeV/$c$.  To reduce the
background from processes that do not contain $W \rightarrow \ell\nu$
decays, we require a transverse mass~\cite{bib:smithUA1} of
$M_T^{\ell\nu} >30$~GeV/$c^2$.  In addition, we restrict
$M_T^{\mu\nu} < 200$~GeV/$c^2$ to suppress muon candidates with poorly
measured momenta.  Candidate events in the electron channel are
required to satisfy a single electron trigger or a trigger requiring
electrons and jets, which results in a combined trigger efficiency for 
the $e\nu{jj}$ selection of $(98^{+2}_{-3})\%$.  A suite of triggers in the 
muon channel achieves a trigger efficiency of $(95\pm5)\%$ for 
the $\mu\nu{jj}$ selection.  Lepton candidates must be spatially matched 
to a track that originates from the $p\bar{p}$ interaction vertex and 
they must be isolated from other energy depositions in the calorimeter 
and other tracks in the central tracking detector.

Most background processes are modeled using Monte Carlo (MC) simulation 
as in the CDF analysis.  Diboson contributions ($WW$, $WZ$, $ZZ$) are 
generated with {\sc pythia}~\cite{bib:PYTHIA} using \textsc{CTEQ6L1} 
parton distribution functions (PDF)~\cite{bib:CTEQ}.  The fixed-order 
matrix element (FOME) generator {\sc alpgen}~\cite{bib:ALPGEN} with
\textsc{CTEQ6L1} PDF is used to generate $W$+jets, $Z$+jets, and $t\bar{t}$
events.  The FOME generator {\sc comphep}~\cite{bib:CompHEP} is used
to produce single top-quark MC samples with \textsc{CTEQ6M} PDF.  Both
{\sc alpgen} and {\sc comphep} are interfaced to {\sc pythia} for
subsequent parton showering and hadronization.  The MC events
undergo a {\sc geant}-based~\cite{bib:GEANT} detector simulation and
are reconstructed using the same algorithms as used for D0 data.  The 
effect of multiple $p\bar{p}$ interactions is included by overlaying data 
events from random beam crossings on simulated events.  All MC samples 
except the $W$+jets are normalized to next-to-leading order (NLO) or 
next-to-NLO (NNLO) predictions for SM cross sections;  
the $t\bar{t}$, single $t$, and diboson cross sections are taken from 
Ref.~\cite{bib:xsecsTT}, Ref.~\cite{bib:xsecsT}, and the \textsc{MCFM} 
program~\cite{bib:mcfm}, respectively.  The $Z$+jets sample is 
normalized to the NNLO cross section~\cite{bib:xsecsZ}.  The multijet background, 
in which a jet misidentified as an isolated lepton passes all selection 
requirements, is determined from data.  In the muon channel, the multijet 
background is modeled with data events that fail the muon isolation 
requirements, but pass all other selections.  In the electron channel, 
the multijet background is estimated using a data sample containing 
events that pass loosened electron quality requirements, but fail the 
tight electron quality criteria.  All multijet samples are corrected 
for contributions from processes modeled by MC.  The multijet
normalizations in the two lepton channels are determined from fits to
the $M_T^{\ell\nu}$ distributions, in which the multijet and $W$+jets
relative normalizations are allowed to float.  The expected rate of
multijet background is determined by this normalization, with an
assigned uncertainty of $20\%$.

Corrections are applied to the MC to account for differences from data
in reconstruction and identification efficiencies of leptons and
jets.  Also, trigger efficiencies measured in data are applied to MC.
The instantaneous luminosity profile and 
$z$ position of the $p\bar{p}$ interaction vertex of each MC sample 
are adjusted to match those in data.  The $p_T$ distribution of $Z$ 
bosons is corrected at the generator level to reproduce dedicated 
measurements~\cite{bib:zptrew}.

Other D0 analyses of this final state apply additional corrections to
improve the modeling of the $W$+jets and $Z$+jets production in the
MC~\cite{bib:whlvbbD0}.  For the results presented in this Letter, we
choose not to apply those corrections in order to parallel the CDF
analysis.  We did, however, study the effects of applying such
corrections~\cite{bib:epaps} and find they do not alter our conclusions.

We consider the effect of systematic uncertainties on both the
normalization and the shape of dijet invariant mass distributions.
Systematic effects are considered from a range of sources: the choice
of renormalization and factorization scales, the {\sc alpgen}
parton-jet matching algorithm~\cite{bib:mlm}, jet energy resolution,
jet energy scale, and modeling of the underlying event and parton showering.
Uncertainties on the choice of PDF, as well as uncertainties from
object reconstruction and identification, are evaluated for all MC
samples.

\begin{figure}[tbp]
  \begin{centering}
    \includegraphics[width=3.3in]{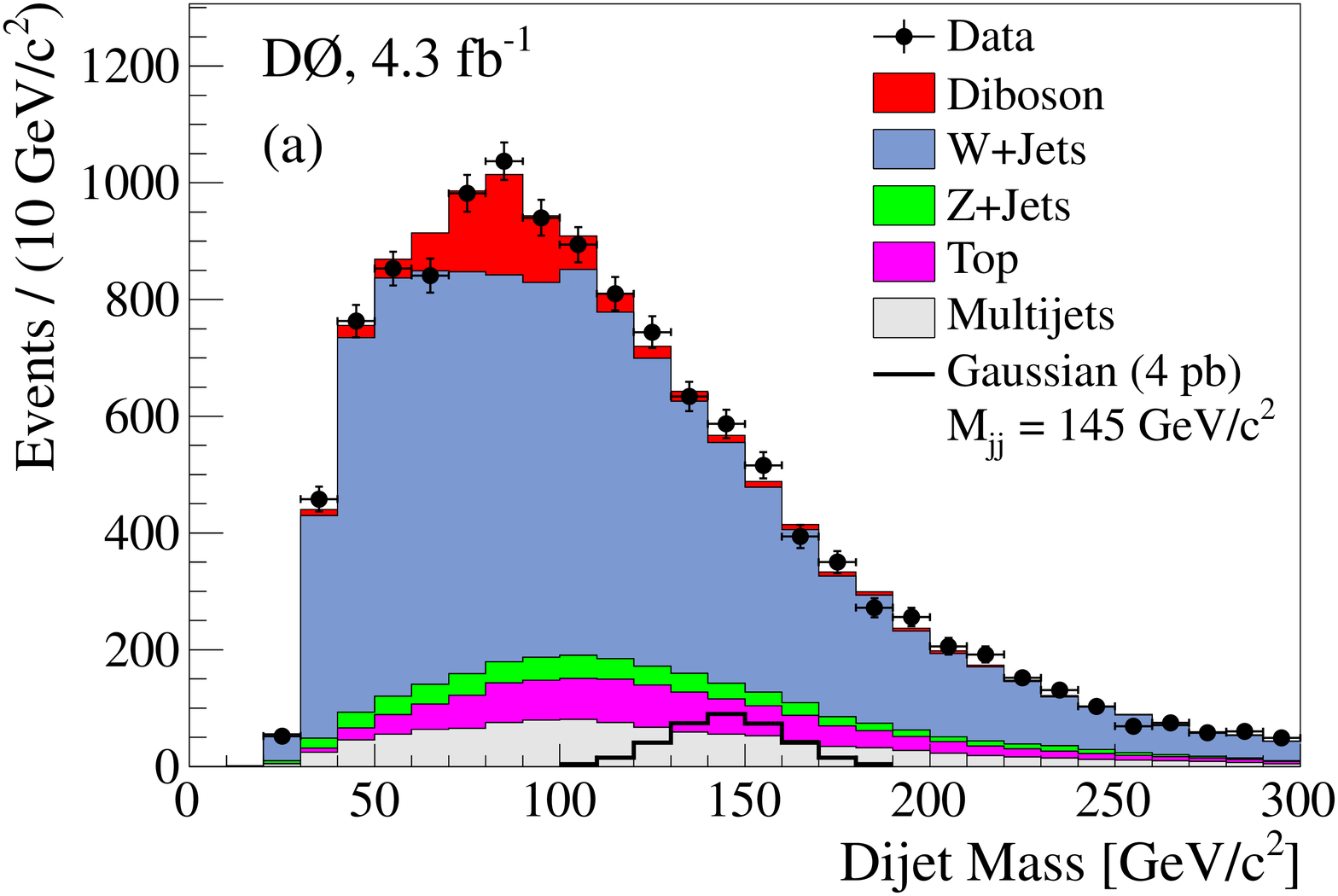}\\
    \includegraphics[width=3.3in]{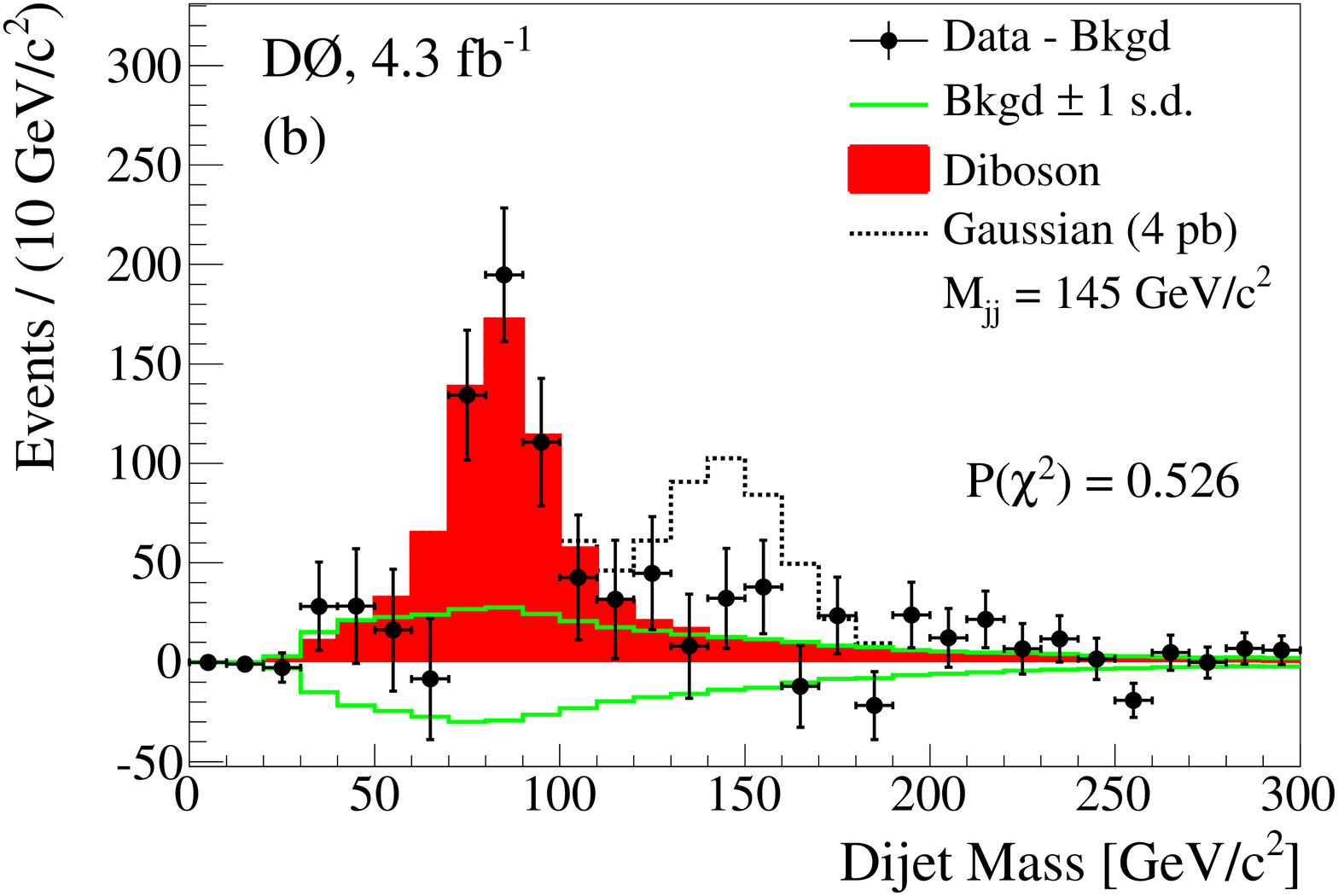} 
    \caption{(color online) Dijet invariant mass summed over electron 
    and muon channels after the fit without (a) and with (b) subtraction of
    SM contributions other than that from the SM diboson processes, along
    with the $\pm$1~s.d.~systematic uncertainty on all SM predictions. 
    The $\chi^{2}$ fit probability, P$(\chi^{2})$, is based on the residuals 
    using data and MC statistical uncertainties. Also shown is the 
    relative size and shape for a model with a Gaussian resonance with a 
    production cross section of 4~pb at $M_{jj}=145$~GeV/$c^2$.  }
    \label{fig:Fig1}
  \end{centering}
\end{figure}

In Fig.~\ref{fig:Fig1} we present the dijet invariant mass
distribution after a fit of the sum of SM contributions to data.
Other distributions are available in the supplementary
material~\cite{bib:epaps}.  The fit minimizes a Poisson
$\chi^2$-function with respect to variations in the rates of
individual background sources and systematic uncertainties that
may modify the predicted dijet invariant mass
distribution~\cite{bib:pflh}.  A Gaussian prior is used for each
systematic uncertainty, including those on the normalization of each 
sample, but the cross sections for diboson and
$W$+jets production in the MC are floated with no constraint.  The 
fit computes the optimal values of the systematic uncertainties,
accounting for departures from the nominal predictions by including a
term in the fit function that sums the squared deviation of each
systematic in units normalized by its $\pm1$~s.d.  Different 
uncertainties are assumed to be mutually
independent, but those common to both lepton channels are treated as
fully correlated.  We perform fits to electron and muon selections
simultaneously and then sum them to obtain the dijet invariant mass
distributions shown in Fig.~\ref{fig:Fig1}.  The measured yields after
the fit are given in Table~\ref{tab:yields}.

\begin{table}[tbp]
  \caption{Yields determined following a $\chi^2$ fit to the data, 
  as shown in Fig.~\ref{fig:Fig1}.  The total uncertainty
  includes the effect of correlations between the individual
  contributions as determined using the covariance matrix.}
  \label{tab:yields}
  \begin{ruledtabular}
    \begin{tabular}{l @{\extracolsep{\fill}} r @{\extracolsep{\fill}} r @{\extracolsep{\fill}} r @{$\ \pm\ $\extracolsep{0cm}} l @{\extracolsep{\fill}} r @{\extracolsep{\fill}} r @{$\ \pm\ $\extracolsep{0cm}} l}
                   & \multicolumn{4}{c}{Electron channel} & \multicolumn{3}{c}{Muon channel} \\
        \hline
        Dibosons                 &&&  434  &   38  &&  304  &  25\\
        $W$+jets                 &&& 5620  &  500  && 3850  & 290\\
        $Z$+jets                 &&&  180  &   42  &&  350  &  60\\
        $t\bar{t}$ + single top  &&&  600  &   69  &&  363  &  39\\
        Multijet                 &&&  932  &  230  &&  151  &  69\\
        \hline
        Total predicted          &&& 7770  &  170  && 5020  & 130\\
        Data                     & \multicolumn{4}{c}{7763} & \multicolumn{3}{c}{5026} \\
    \end{tabular}
  \end{ruledtabular}
\end{table}

To probe for an excess similar to that observed by the CDF
Collaboration~\cite{bib:CDFbump}, we model a possible signal as a 
Gaussian resonance in the dijet invariant mass with an observed 
width corresponding to the expected resolution of the D0 detector given by 
$\sigma_{jj} = \sigma_{W \to jj} \cdot \sqrt{M_{jj}/M_{W\to
jj}}$. Here, $\sigma_{W\to jj}$ and $M_{W \to jj}$ are the width
and mass of the $W \to jj$ resonance, determined to be $\sigma_{W \to
jj} =11.7$~GeV/$c^2$ and $M_{W\to jj} =81$~GeV/$c^2$ from a simulation 
of $WW \to \ell\nu jj$ production.
For a dijet invariant mass resonance at $M_{jj} =145$~GeV/$c^2$, the 
expected width is $\sigma_{jj} =15.7$~GeV/$c^2$.

We normalize the Gaussian model in the same way as
reported in the CDF Letter~\cite{bib:CDFbump}.  We assume that any such excess comes from
a particle $X$ that decays to jets with 100\% branching fraction.  The
acceptance for this hypothetical process ($WX \to \ell\nu jj$) is estimated 
from a MC simulation of $WH \to \ell \nu b\bar{b}$ production.  When testing 
the Gaussian signal with a mean of $M_{jj} =145$~GeV/$c^2$, the acceptance
is taken from the $WH \to \ell \nu b\bar{b}$ simulation with $M_H
=150$~GeV/$c^2$.  This prescription is chosen to be consistent with
the CDF analysis, which used a simulation of $WH\to \ell \nu b\bar{b}$ 
production with $M_H =150$~GeV/$c^2$ to estimate the acceptance for the 
excess that they observes at $M_{jj}=144$~GeV/$c^2$.  When probing other values 
of $M_{jj}$, we use the acceptance obtained for $WH\to\ell\nu b\bar{b}$ MC 
events with $M_H =M_{jj} +5$~GeV/$c^2$.

We use this Gaussian model to derive upper limits on the cross
section for a possible dijet resonance as a function of dijet
invariant mass using the $CL_s$ method with a negative log-likelihood
ratio (LLR) test statistic~\cite{bib:cls1} that is summed over all
bins in the dijet invariant mass spectrum. Upper limits on cross
section are calculated at the 95\% confidence level (C.L.) for
Gaussian signals with mean dijet invariant mass in the range
110~$<M_{jj}<170$~GeV/$c^2$, in steps of 5~GeV/$c^2$, allowing the
cross sections for $W$+jets production to float with no
constraint. Other contributions are constrained by the {\it a priori}
uncertainties on their rate, either derived from theory or subsidiary
measurements.

\begin{figure}[tbp]
  \begin{centering}
    \includegraphics[width=3.3in]{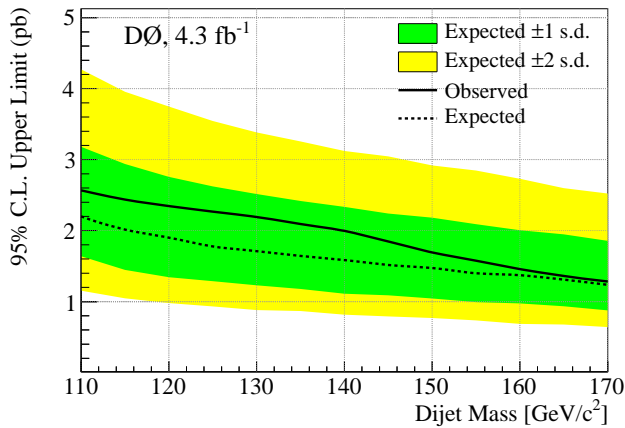}
    \caption{(color online) Upper limits on the cross section (in pb) 
    at the 95\%~C.L.  for a Gaussian signal in dijet invariant mass.  
    Shown are the limit expected using the background prediction, the 
    observed data, and the regions corresponding to a 1~s.d.~and 
    2~s.d.~fluctuation of the backgrounds.}
    \label{fig:Fig2} 
  \end{centering}
\end{figure}

The Gaussian model is assigned systematic uncertainties
affecting both the normalization and shape of the distribution derived
from the systematic uncertainties on the diboson simulation.  A
fit~\cite{bib:pflh} of both the signal+background and
background-only hypotheses is performed for an ensemble of
pseudo-experiments as well as for the data distribution.  The results
of the cross section upper limit calculation are shown in
Fig.~\ref{fig:Fig2} and are summarized in Table~\ref{tab:limits}.

\begin{table*}[htpb]
  \begin{ruledtabular}
    \caption{Expected and observed upper limits on the cross section
    (in pb) at the 95\%~C.L. for a dijet invariant mass resonance.}
    \label{tab:limits}
    \begin{tabular}{lccccccccccccc}
$M_{jj}$~(GeV) &110 &115 &120 &125 &130 &135 &140 &145 &150 &155 &160 &165 &170\\
\hline
Expected: &2.20 &2.01 &1.90 &1.78 &1.71 &1.64 &1.58 &1.52 &1.47 &1.40 &1.37 &1.31 &1.24 \\
Observed: &2.57 &2.44 &2.35 &2.27 &2.19 &2.09 &2.00 &1.85 &1.69 &1.58 &1.46 &1.36 &1.28 \\
    \end{tabular}
  \end{ruledtabular} 
\end{table*}

In a further effort to evaluate the sensitivity for any excess of 
events of the type reported by the CDF Collaboration, we perform a 
signal-injection test.  We repeat the statistical analysis after 
injecting a Gaussian signal model, normalized to a cross section 
of 4~pb, into the D0 data sample, thereby creating a mock ``data'' 
sample modeling the expected outcome with a signal present.  The 
size and shape of the injected Gaussian model for 
$M_{jj}=145$~GeV/$c^2$ relative to other data components is shown 
in Fig.~\ref{fig:Fig1}.

\begin{figure}[tbp]
  \begin{centering}
    \includegraphics[width=3.3in]{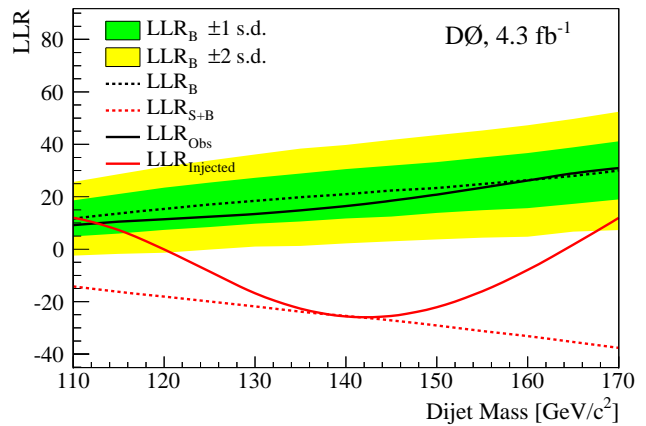}\\
    \caption{(color online) Log-likelihood ratio test statistic as a function of
    probed dijet mass.  Shown are the expected LLR for the background
    prediction (dashed black) with regions corresponding to a
    1~s.d.~and 2~s.d.~fluctuation of the backgrounds, for the
    signal+background prediction (dashed red), for the observed
    data (solid black), and for data with a dijet
    invariant mass resonance at 145~GeV/$c^2$ injected with a cross
    section of 4~pb (solid red).}
    \label{fig:Fig3} 
  \end{centering} 
\end{figure}

The LLR metric provides a sensitive measure of model compatibility,
providing information on both the rate and mass of any signal-like
excess.  We therefore study the LLR distributions obtained with actual
data as well as the signal-injected mock data sample.  The results of
the LLR test in Fig.~\ref{fig:Fig3} show a striking difference between
the two hypotheses, demonstrating that this analysis is sensitive to
the purported excess.  In the actual data, however, no significant
evidence for an excess is observed.  

\begin{figure}[tbp]
  \begin{centering}
    \includegraphics[width=3.3in]{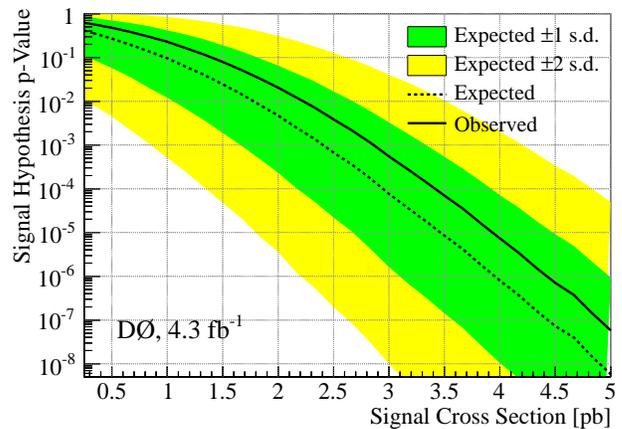}\\
    \caption{(color online) Distribution of $p$-values for the signal+background
    hypothesis with a Gaussian signal with mean of
    $M_{jj} =145$~GeV/$c^2$ as a function of hypothetical signal
    cross section (in pb).  Shown are the $p$-values for the background
    prediction (dashed black) with regions corresponding to a 
    1~s.d.~and 2~s.d.~fluctuation of the backgrounds and the observed data 
    (solid black).}
    \label{fig:Fig4} 
  \end{centering} 
\end{figure}

In Fig.~\ref{fig:Fig4}, we show as a function of cross section the 
$p$-value obtained by integrating the LLR distribution populated 
from pseudo-experiments drawn from the signal+background 
hypothesis above the observed LLR, assuming a Gaussian invariant 
mass distribution with a mean of $M_{jj}=145$~GeV/$c^2$.  The 
$p$-value for a Gaussian signal with cross section of 4~pb is 
$8.0 \times 10^{-6}$, corresponding to a rejection of this signal 
cross section at a Gaussian equivalent of 4.3~s.d.  We set a 
95\%~C.L. upper limit of 1.9~pb on the production cross section of 
such a resonance.

In summary, we have used 4.3~fb$^{-1}$ of integrated luminosity
collected with the D0 detector to study the dijet invariant mass
spectrum in events containing one $W\to\ell\nu$ ($\ell=e$ or 
$\mu$) boson decay and two high-$p_T$ jets.  Utilizing a similar 
data selection as the CDF Collaboration we find no evidence for 
anomalous, resonant production of dijets in
the mass range $110 - 170$~GeV/$c^2$.  Using a simulation of $WH \to \ell
\nu b\bar{b}$ production to model acceptance and efficiency, we derive
upper limits on the cross section for anomalous resonant dijet
production.  For $M_{jj} = 145$~GeV/$c^2$, we set a 95\%~C.L. upper
limit of 1.9~pb on the cross section and we reject the hypothesis 
of a production cross section of 4~pb at the level of 4.3~s.d.  
In the case that the cross section reported by the CDF Collaboration 
is modified, we report in Fig.~\ref{fig:Fig4} the variation of our 
$p$-value for exclusion of potential resonance cross sections other 
than 4~pb.

\input acknowledgement.tex
\clearpage

\section{Appendix}

\subsection{Fit of a 145~\boldmath${\rm{GeV}/c^2}$ Dijet Resonance}

\begin{figure}[htb]
    \begin{centering}
      \includegraphics[width=3.3in]{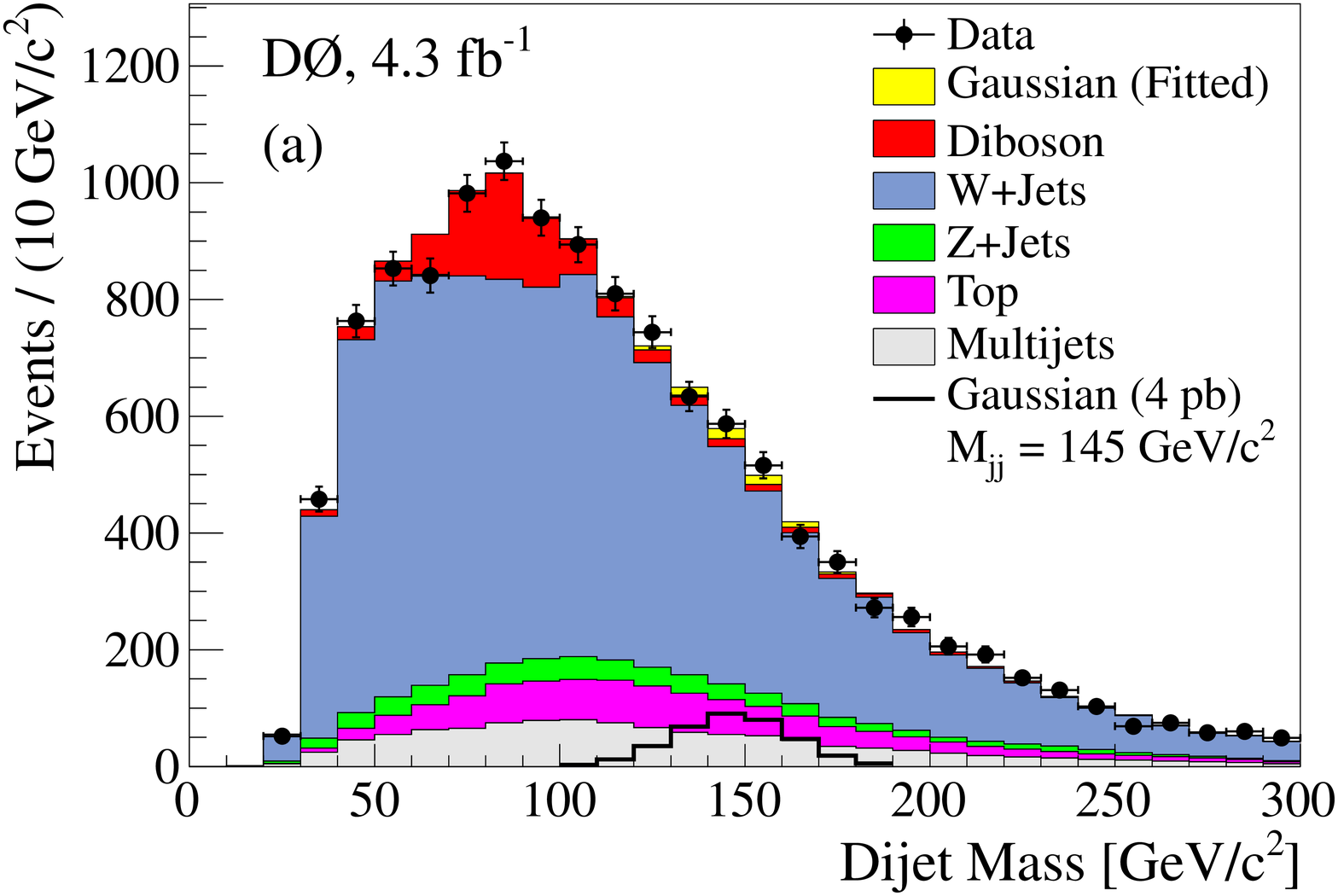}
      \includegraphics[width=3.3in]{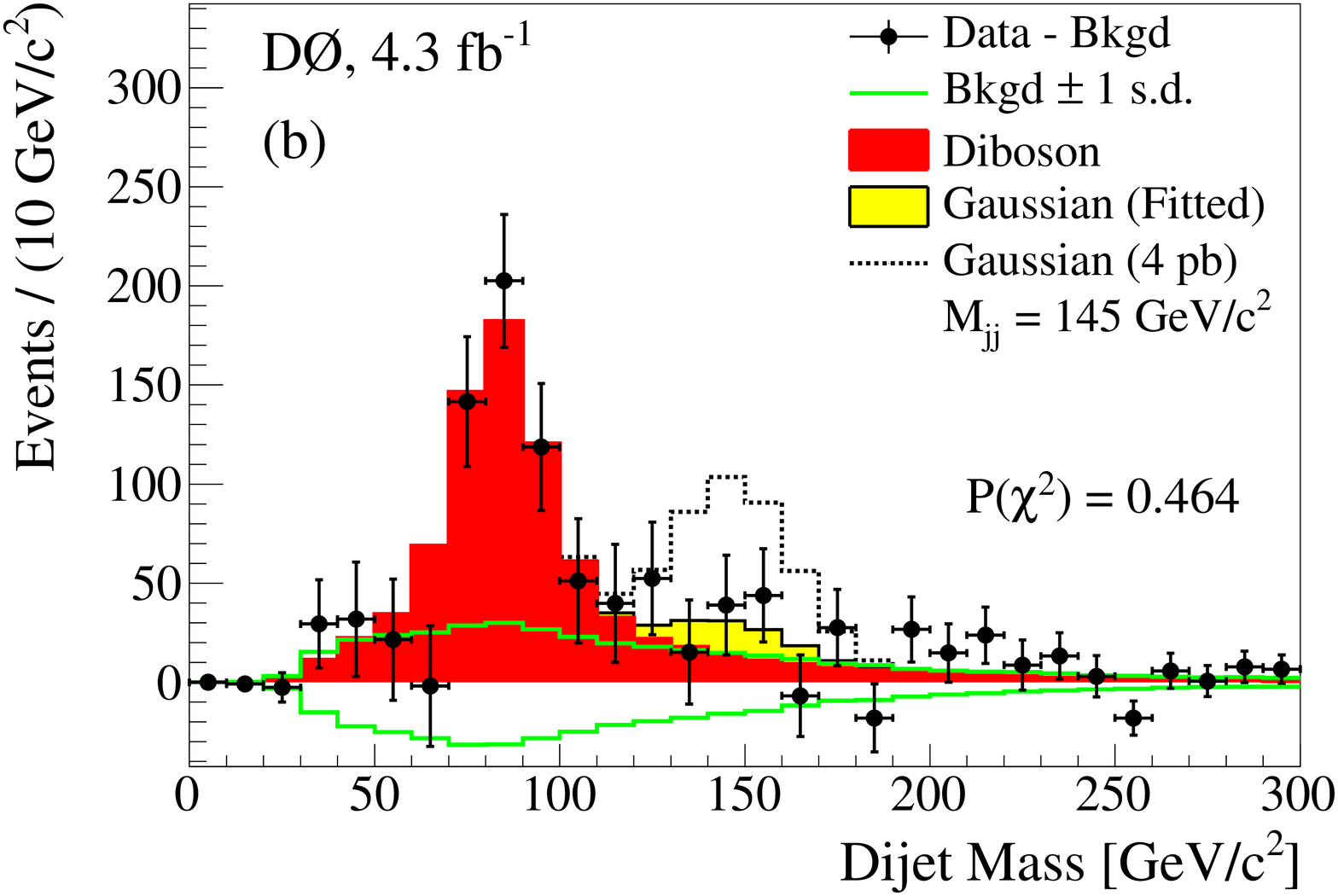}
      \caption{(color online) Dijet invariant mass summed over lepton channels when
	including the Gaussian model in the fit to data (a)
	without and (b) with subtraction of SM contributions other than
	from the SM diboson processes, along with the 
        $\pm$1~s.d.~systematic uncertainty on all SM predictions.  }
      \label{fig:fitWX} 
    \end{centering}
  \end{figure}

  The D0 data do not indicate the presence of a non-SM dijet resonance
  such as indicated by the CDF Collaboration.  In the Letter we
  showed the fit of the SM predictions to the data.  However, fitting
  only the SM contributions could hide an excess if the systematic
  uncertainties allowed the SM contributions to be distorted in such a
  way that they filled in the excess.  To study this question, we
  present a fit to the data of the SM predictions plus the Gaussian
  signal template with $M_{jj} = 145$~GeV/$c^2$.  The resulting dijet
  mass distribution is shown in Fig.~\ref{fig:fitWX}.  The fit is
  performed in the same way as described in the Letter (no
  constraint on diboson or $W$+jets normalizations), except that it
  now also includes the Gaussian signal template 
  for $M_{jj} = 145$~GeV/$c^2$ with a freely floating normalization.  
  The Gaussian model includes
  systematic uncertainties affecting both the normalization and shape
  of the template, analogous to the systematic uncertainties for the
  diboson prediction.  The best fit value for the Gaussian template
  yields a cross section of $\sigma(p\bar{p}\to WX) =
  0.82^{+0.83}_{-0.82}$~pb, consistent with no excess.  When we fix
  the diboson cross section to the SM prediction with a Gaussian prior
  of 7\% on the rate, the best fit value for the Gaussian template
  yields a cross section of $\sigma(p\bar{p}\to WX) =
  0.42^{+0.76}_{-0.42}$~pb.

  \subsection{Kinematic Corrections to the Simulation}

  \begin{figure*}[htb]
    \begin{centering}
      \includegraphics[width=3.3in]{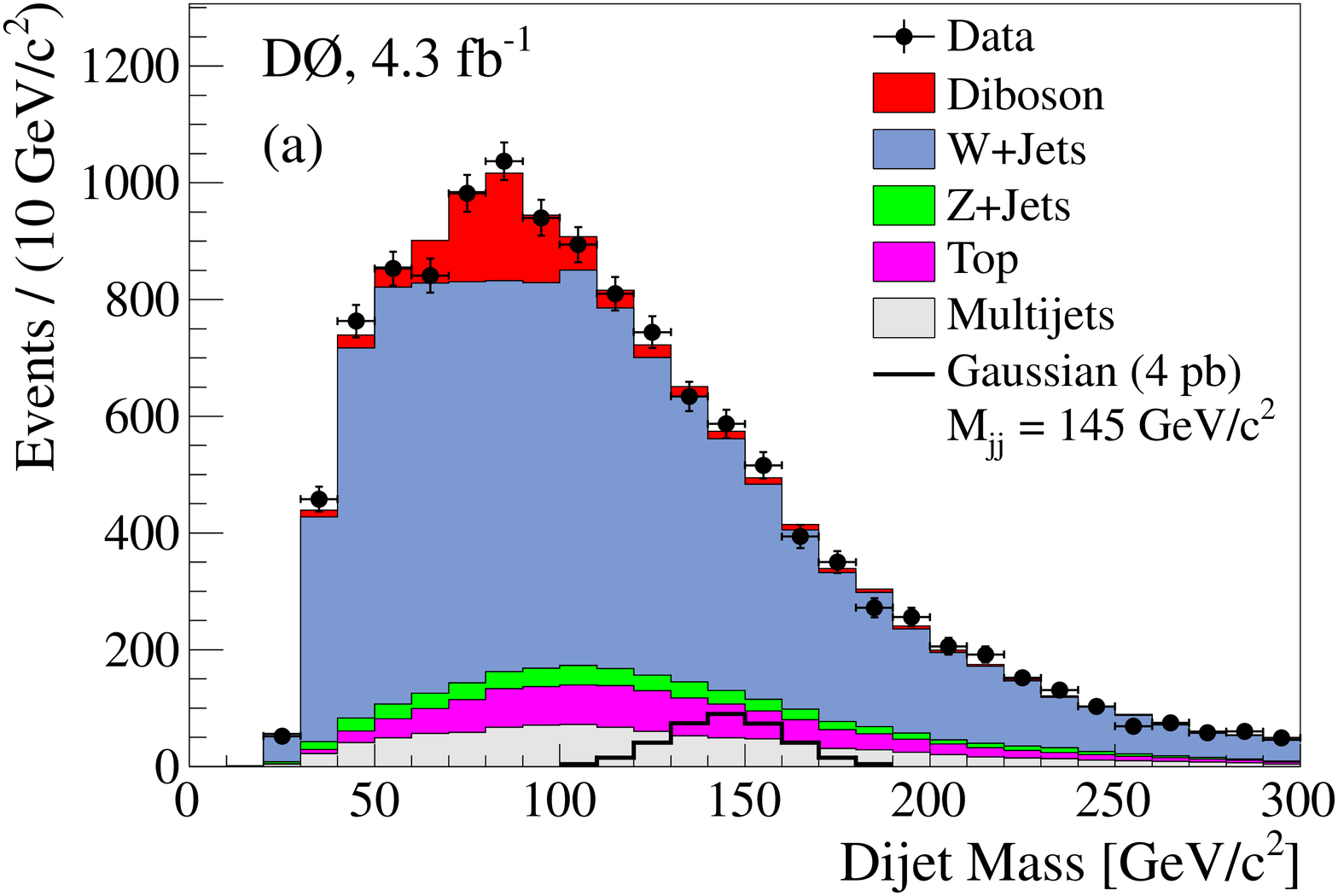}
      \includegraphics[width=3.3in]{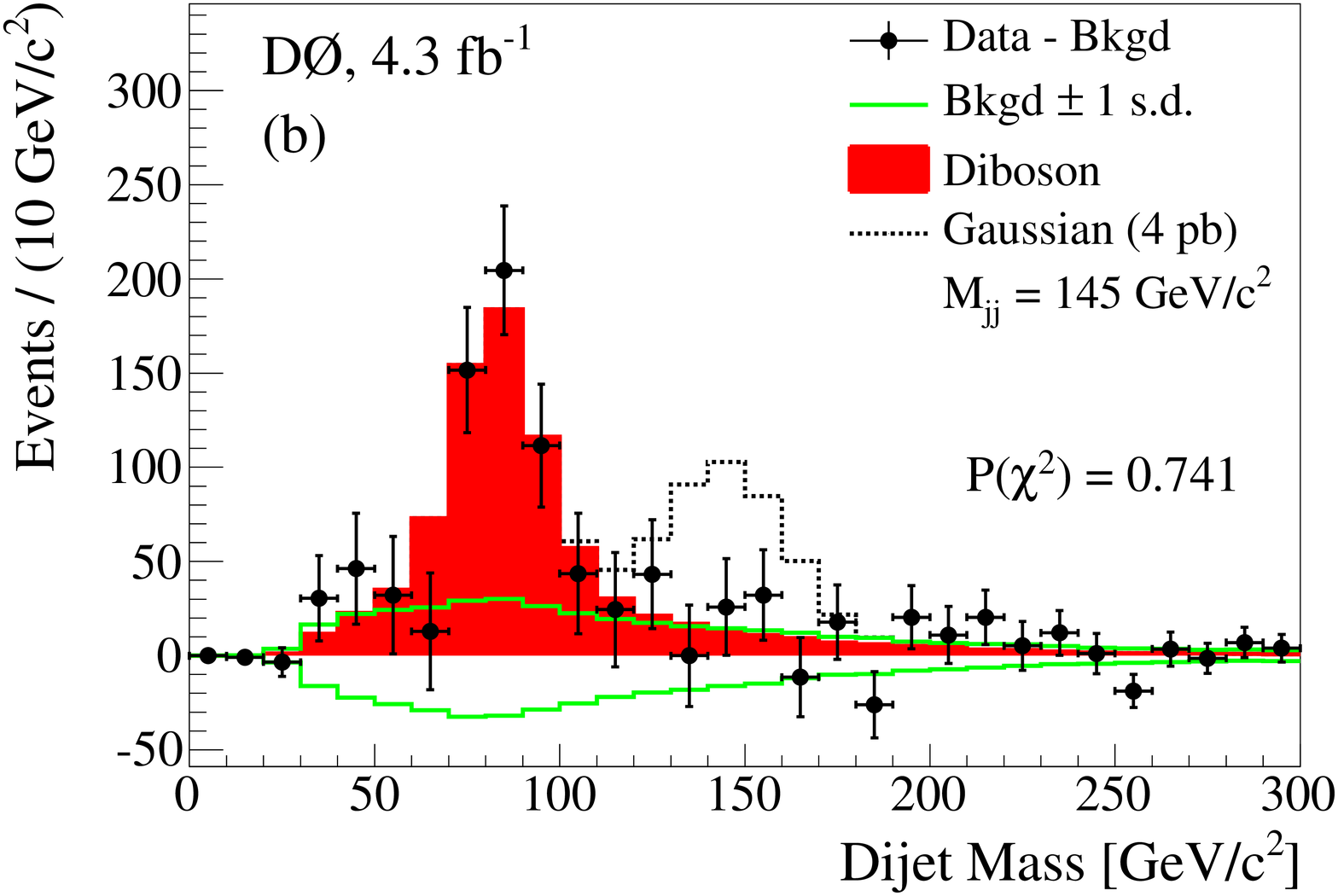}
      \caption{(color online) Dijet invariant mass summed over lepton
	channels after the fit (a) without and (b) with 
        SM contributions subtraction other than from the SM 
        diboson processes, along with the 
        $\pm$1~s.d.~systematic uncertainty on all SM predictions.
	These distributions have the additional kinematic 
        corrections applied to the MC. }
      \label{fig:Fig1RW} 
    \end{centering}
  \end{figure*}

  \begin{table*}[htpb]
    \begin{ruledtabular}
      \caption{Expected and observed upper limits on the cross section (in pb) 
        at the 95\%~C.L. for a dijet invariant mass resonance.  These 
        limits are derived with the additional kinematic corrections 
        applied to the MC.}
      \label{tab:limitsRW}
      \begin{tabular}{lccccccccccccc}
      $M_{jj}$~(GeV) &110 &115 &120 &125 &130 &135 &140 &145 &150 &155 &160 &165 &170 \\
      \hline
 Expected: &2.35 &2.16 &2.05 &1.97 &1.88 &1.81 &1.73 &1.68 &1.65 &1.56 &1.52 &1.45 &1.42 \\
Observed: &2.26 &2.02 &1.93 &1.83 &1.74 &1.64 &1.55 &1.48 &1.37 &1.27 &1.18 &1.09 &1.06 \\
      \end{tabular}
    \end{ruledtabular} 
  \end{table*}

  \begin{figure*}
   \begin{minipage}[htb]{3.3in}
   \begin{center}
    \includegraphics[width=3.3in,clip]{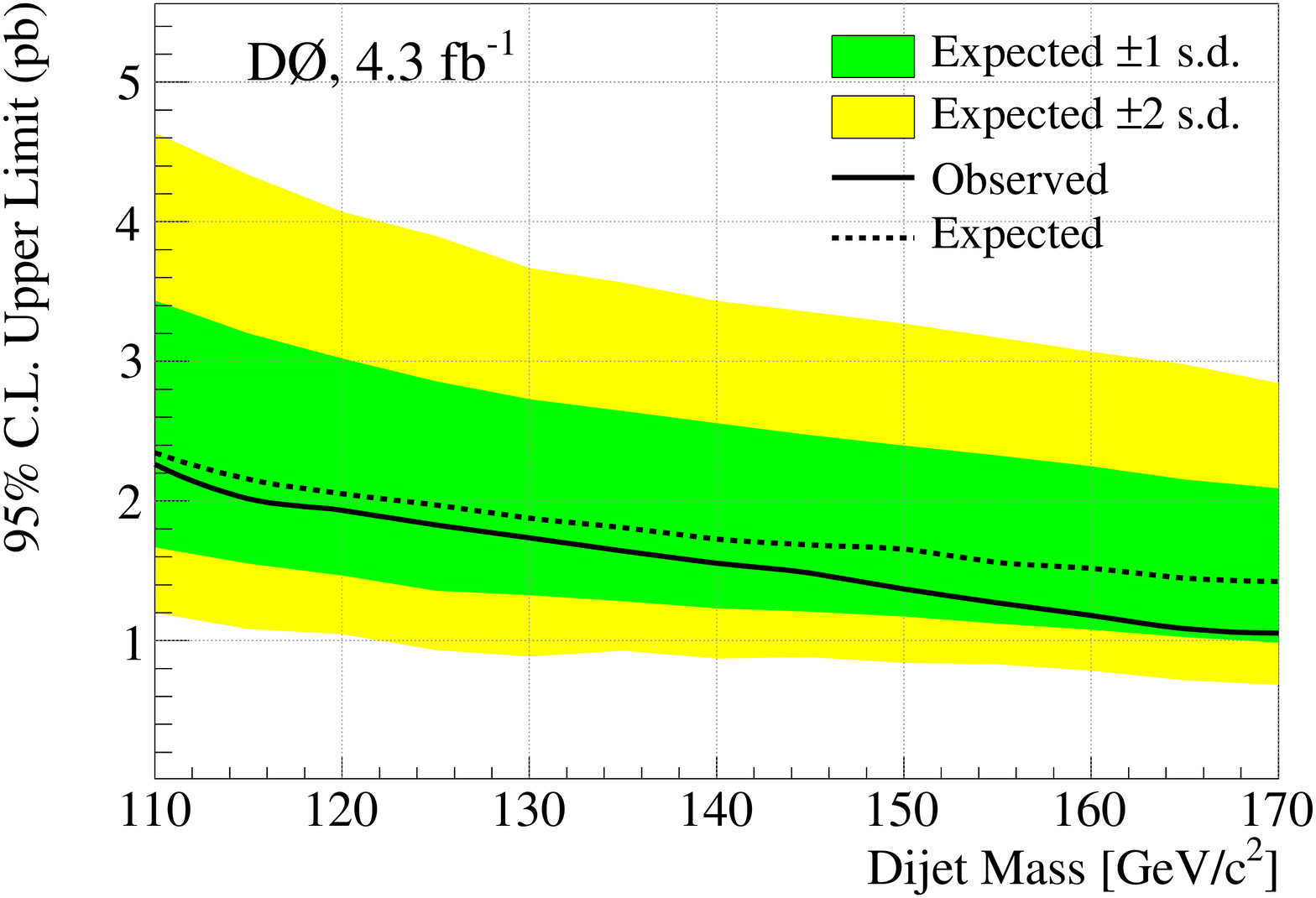}
    \caption{\label{fig:Fig2RW} 
	(color online) Upper limits on the cross section (in pb) at the 95\%~C.L.
	for a Gaussian signal in dijet invariant mass.  These results
	are derived with the additional kinematic corrections applied
	to the MC.  Shown in the figure are the limit expected using
	the background prediction, the observed data limit, and the
	regions corresponding to a 1~s.d.~and 2~s.d.~fluctuation of the
	backgrounds.}
   \end{center}
   \end{minipage} \hfill
   \begin{minipage}[htb]{3.3in}
   \begin{center}
    \includegraphics[width=3.3in,clip]{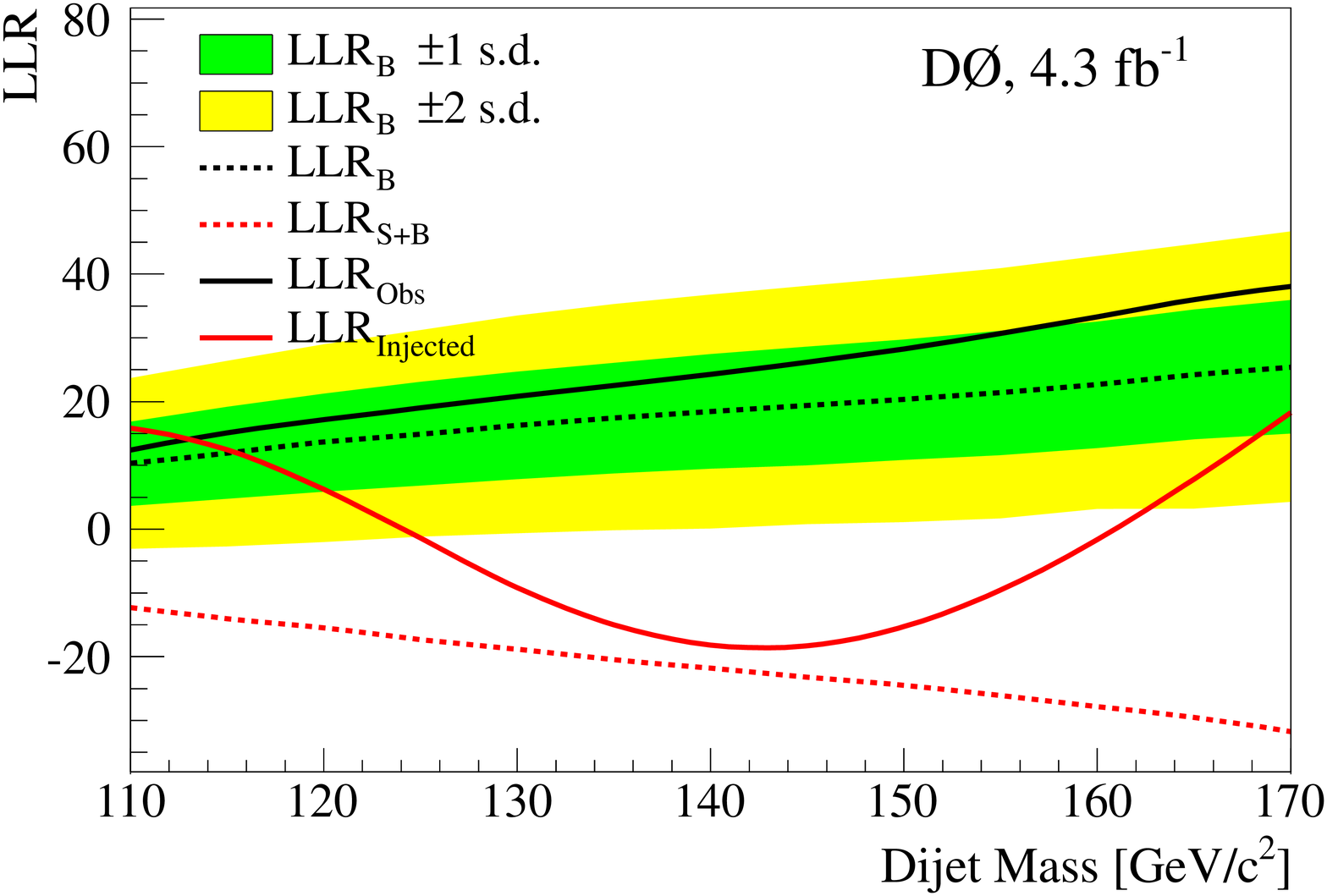}
    \caption{\label{fig:Fig3RW} 
        (color online) Log-likelihood ratio test statistic as a function of
	probed dijet invariant mass.  These results are derived with the
	additional kinematic corrections applied to the MC.  Shown are
	the expected LLR for the background prediction (dashed black)
	with regions corresponding to a 1~s.d.~and 2~s.d.~fluctuation of
	the backgrounds, for the signal+background prediction (dashed
	red), for the data (solid black), and for data with a dijet mass 
        resonance at 145~GeV/$c^2$ injected using a cross section of 4~pb (solid red).}
   \end{center}
   \end{minipage}
   \end{figure*}

  The common tools used to simulate the predicted SM
  contributions perform well in general, but they have shortcomings.  
  For example, different event generators have different predictions 
  for production angles and relative angles between jets in 
  $W$+jets and $Z$+jets events~\cite{bib:ALPGENcomp}.  Thus, it is 
  not unexpected that the simulated $W$+jets and $Z$+jets samples do 
  not perfectly model the angular distributions of jets.  For analyses 
  with looser selections, such as the search for $WH$ production at 
  D0~\cite{bib:whlvbbD0}, these jet angular distributions show clear 
  discrepancies between data and the simulated $W$+jets and $Z$+jets 
  events.  Thus, these analyses use parameterized functions to correct 
  the pseudorapidities of the two highest $p_T$ jets and the 
  $\Delta{\cal R}=\sqrt{(\Delta \eta)^2 + (\Delta \phi)^2}$ separation 
  between those two jets in $W/Z$+jets samples, and the transverse 
  momentum of the $W$ boson candidate, $p_T(W)$, in the $W$+jets samples, 
  to better model the data.

  The tight kinematic selection criteria employed in this analysis
  (e.g., $p_T(jj) > 40~\rm{GeV}/c^2$) remove much of the phase space
  in which the MC generators have difficulty modeling data (e.g., low
  $p_T(W)$), greatly reducing the need for the kinematic corrections
  of the simulation.  Therefore, the plots and results in the Letter 
  do {\bf not} use any of these kinematic corrections, which
  is consistent with the CDF analysis.

  Although kinematic corrections are not required to achieve adequate
  modeling when applying the tight selection criteria of this
  analysis, modeling issues are probably still present.  In this
  section we present the results obtained when the kinematic
  corrections (derived from a selection similar to the search for $WH$
  production~\cite{bib:whlvbbD0}) are applied to this analysis.

  The following figures are analogous to those in the Letter,
  except that the above mentioned kinematic corrections have been
  applied to the simulation.  Figure~\ref{fig:Fig1RW} shows the dijet
  invariant mass distribution after the fit of the sum of SM
  predictions to data.  The change relative to Fig.~1 in the Letter 
  is not large, but improved modeling is evident in the higher
  $\chi^2$ probability.  The resulting upper limits on the cross
  section for production of a dijet invariant mass resonance are
  presented in Table~\ref{tab:limitsRW} and shown in
  Fig.~\ref{fig:Fig2RW}.  They are consistent with those in Fig.~2
  from the Letter.  Figure~\ref{fig:Fig3RW} shows the LLR
  distributions analogous to Fig.~3 from the Letter.
  
  From this study we conclude that the kinematic corrections to the
  simulation do not substantially change the result and we reach the
  same conclusion that there is no excess of dijet events in the D0
  data similar to that reported by the CDF collaboration.

  \subsection{Additional Data-MC Comparisons}
  
\begin{figure*}[htb]
    \begin{centering}
      \includegraphics[width=3.3in]{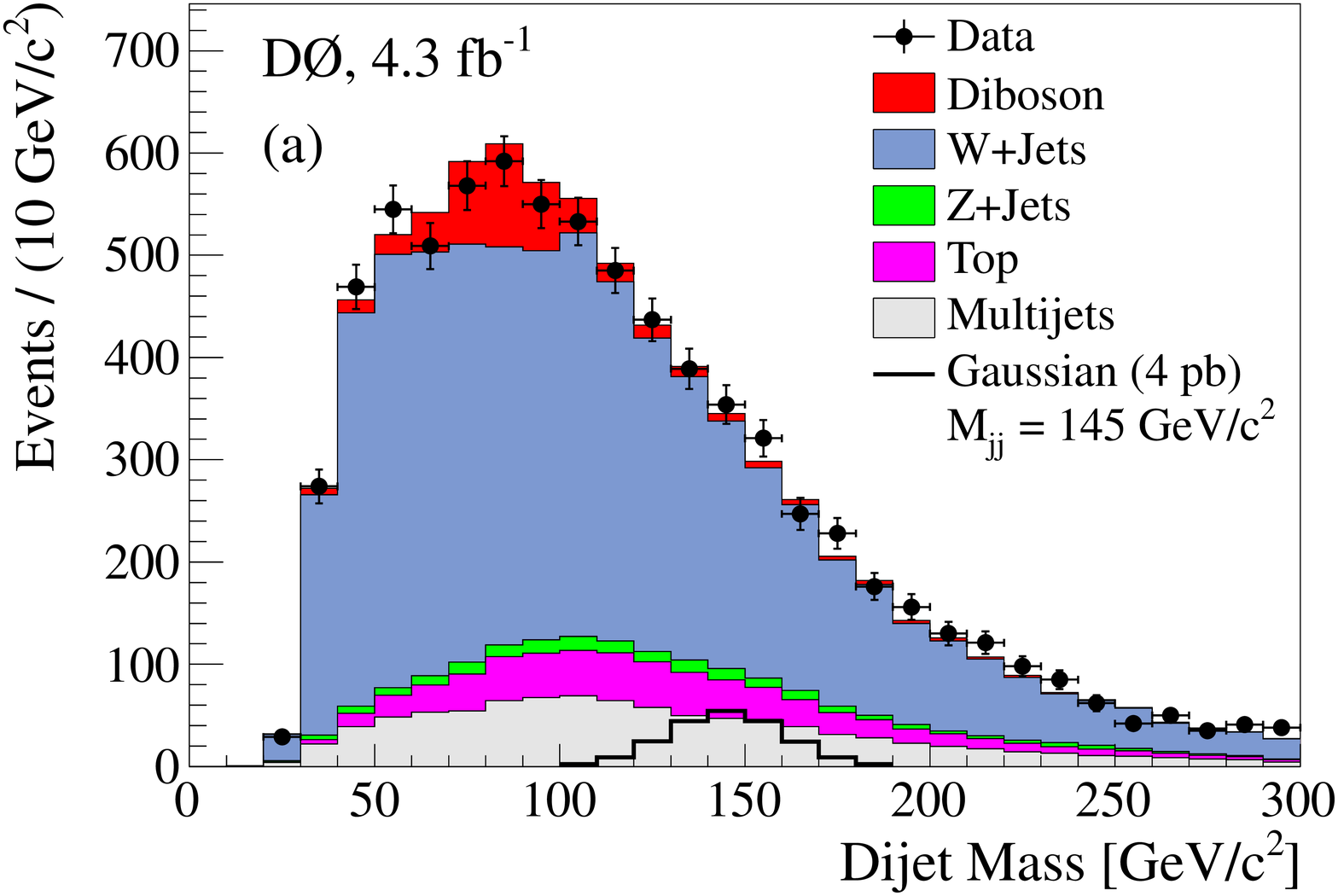}
      \includegraphics[width=3.3in]{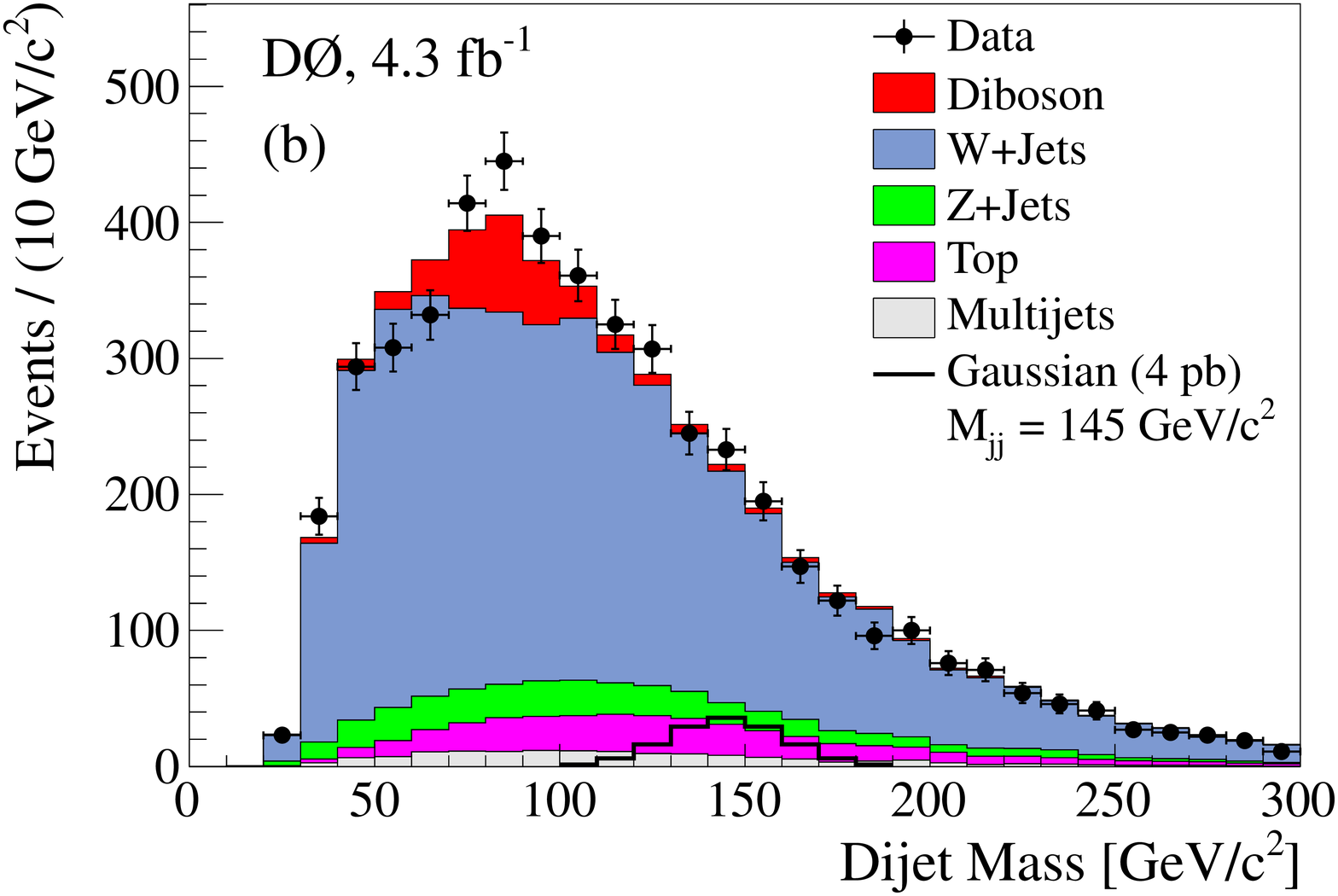} \\
      \caption{(color online) Dijet invariant mass distributions separately for the
      (a) electron and (b) muon channel after the simultaneous fit of
      these two distributions.}
      \label{fig:Separate} 
    \end{centering}
  \end{figure*}

\begin{figure*}[htb]
    \begin{centering}
	\includegraphics[width=3.3in]{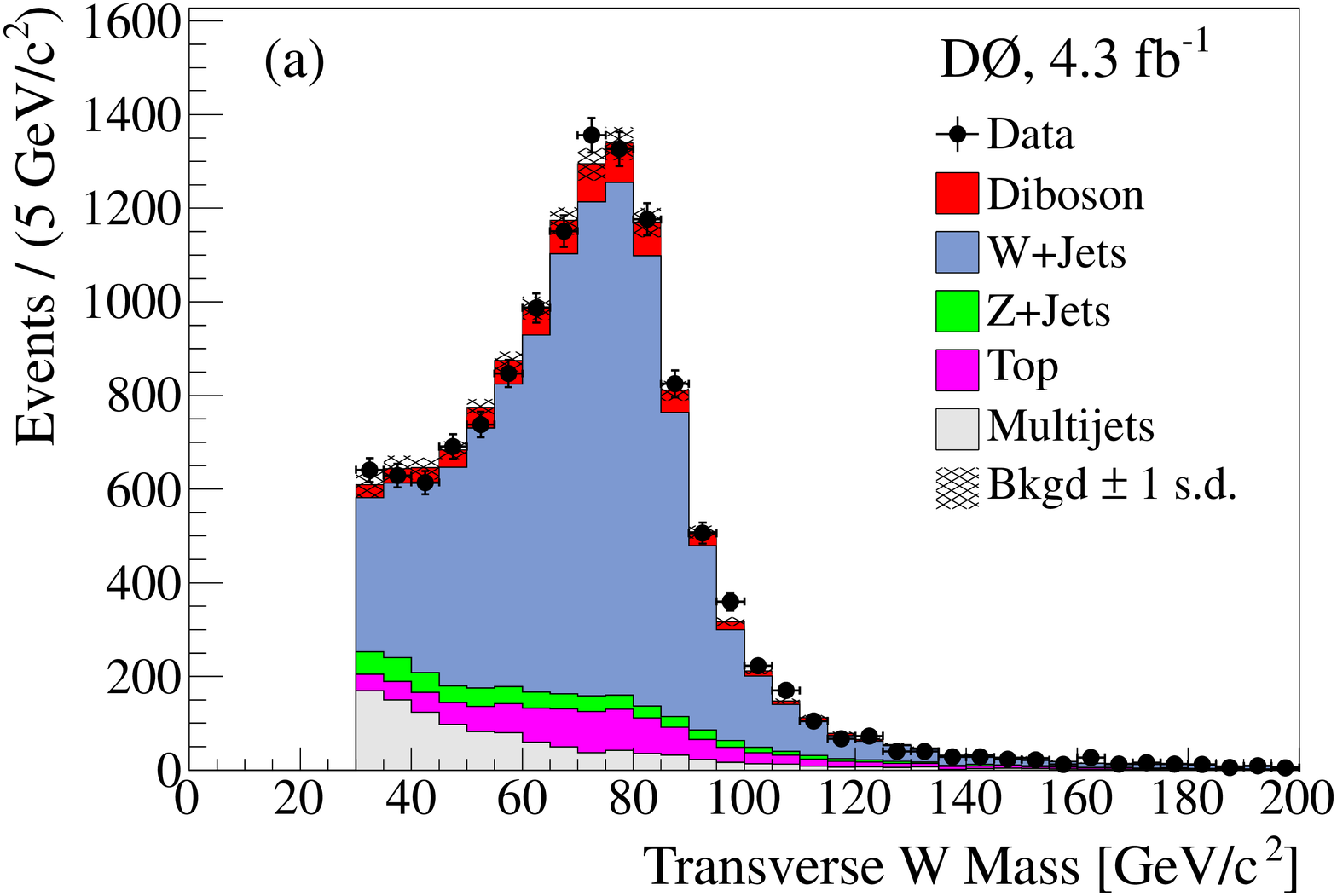}
	\includegraphics[width=3.3in]{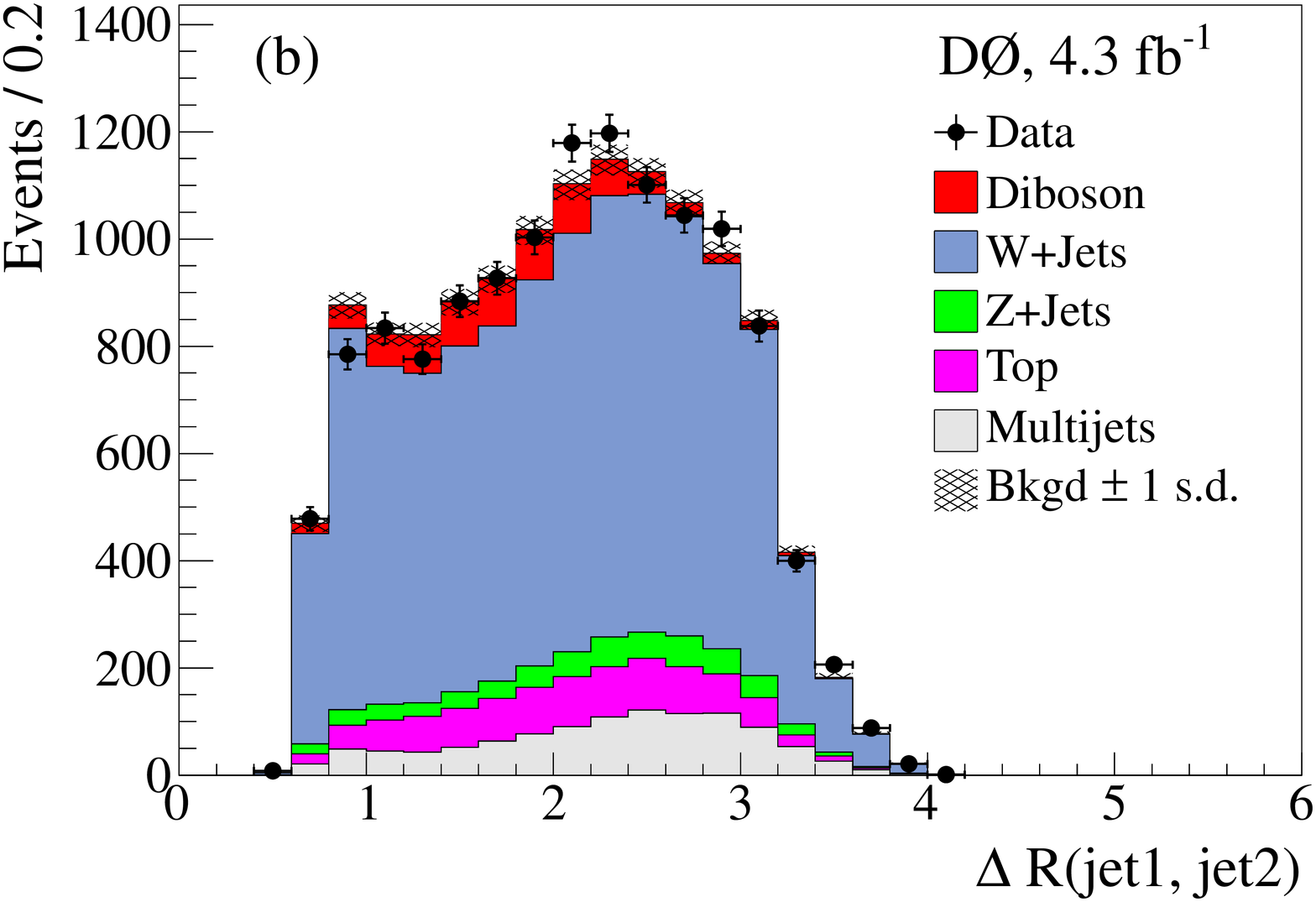} \\
	\includegraphics[width=3.3in]{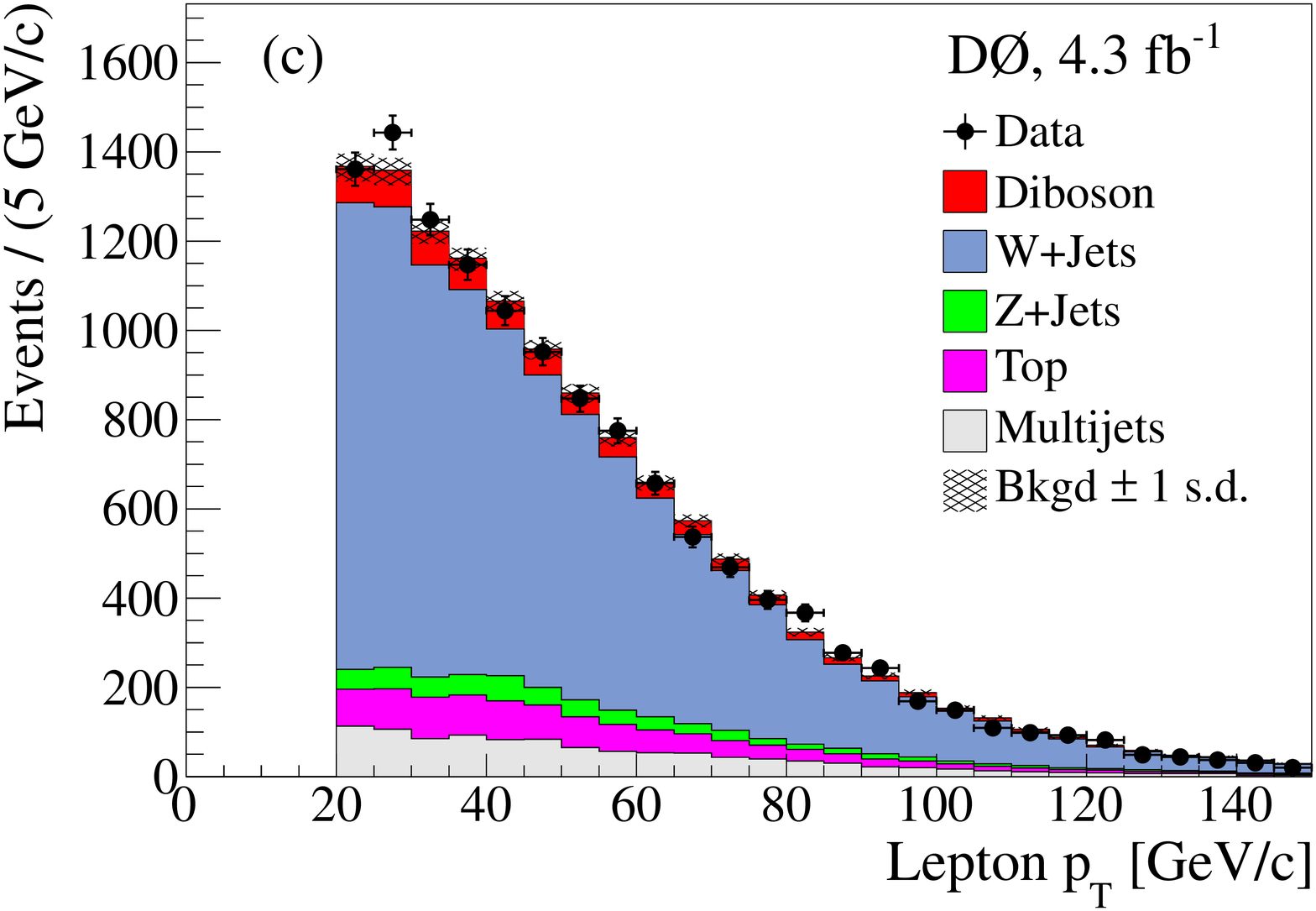}
	\includegraphics[width=3.3in]{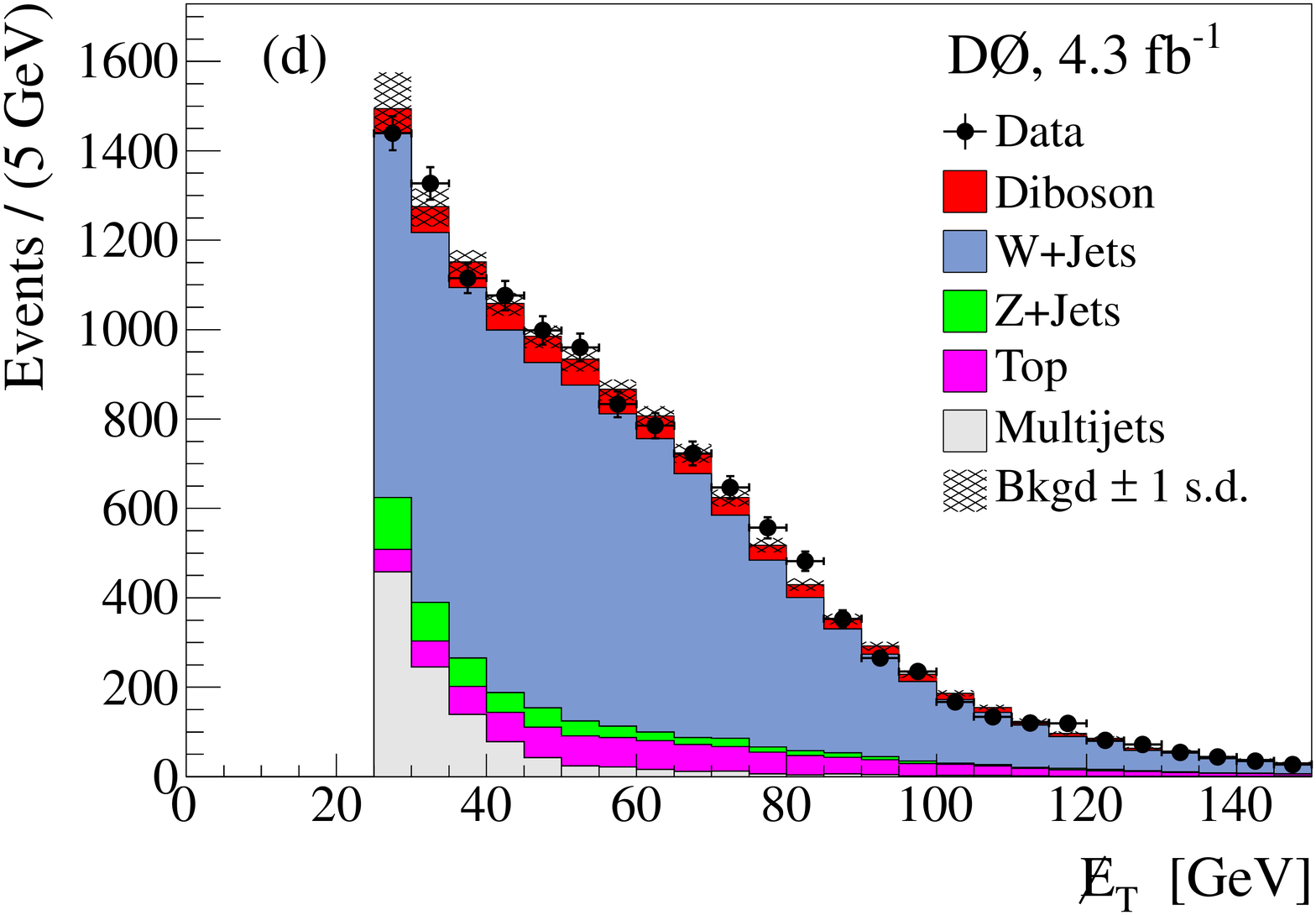} \\
	\includegraphics[width=3.3in]{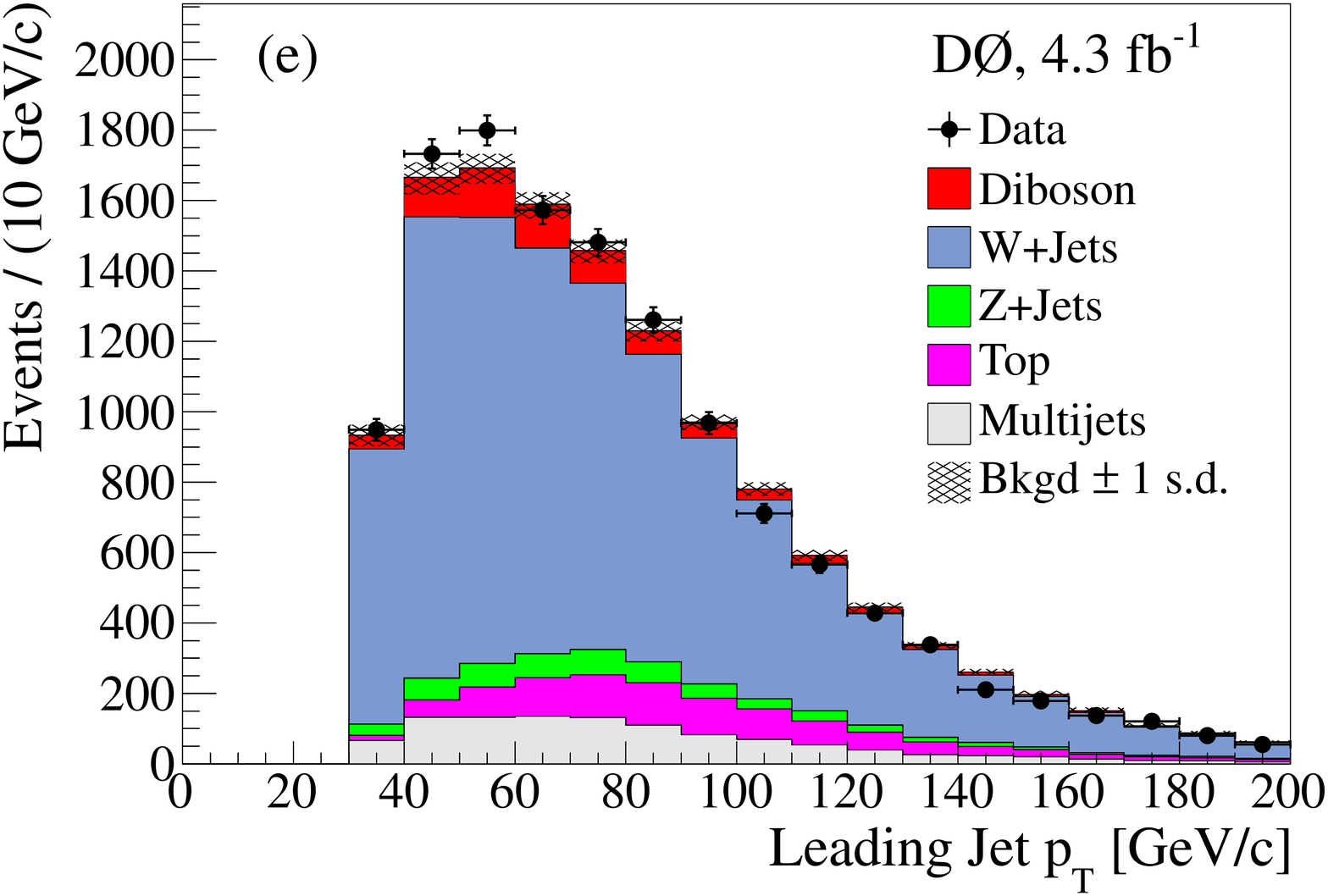}
	\includegraphics[width=3.3in]{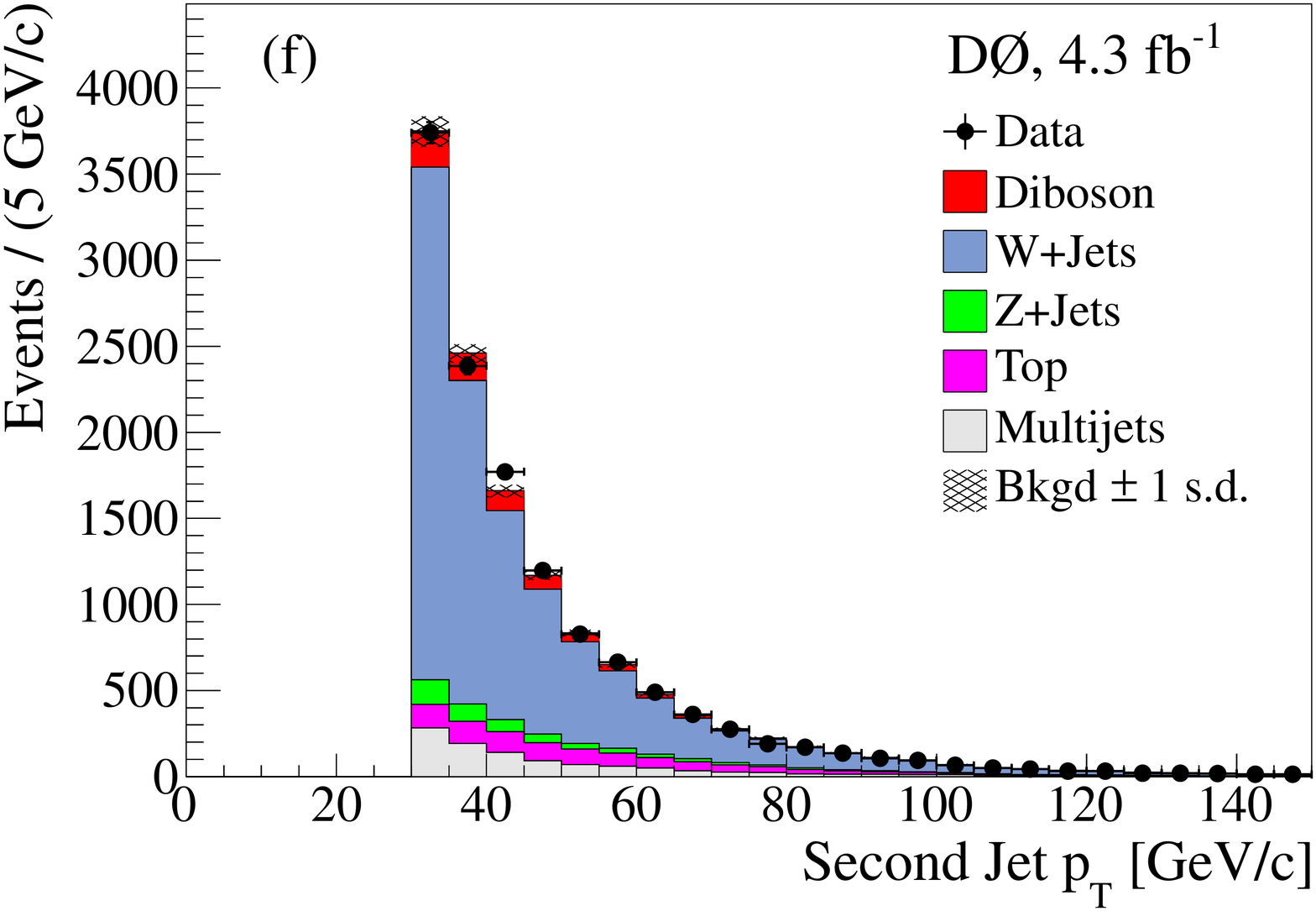} \\
      \caption{(color online) Distributions of kinematic variables (combined electron
	and muon channels) evaluated using the results of a $\chi^2$
	fit of SM predictions to data for the dijet invariant mass
	distribution: (a) transverse $W$ mass, (b) $\Delta{\cal R}$
	separation between the two selected jets, (c) lepton $p_T$,
	(d) missing transverse energy, (e) highest jet $p_T$, (f)
	second highest jet $p_T$. 
        The $\pm$1~s.d.~systematic uncertainty on all SM predictions 
        is presented by the cross-hatched area. } \label{fig:MorePlots}
	\end{centering} \end{figure*}

  Kinematic distributions presented in this Section are modeled without 
  the additional corrections applied to the MC. 
  Figure~\ref{fig:Separate} shows the dijet invariant mass
  distribution for the separate lepton channels after the simultaneous
  fit in these two distributions of the SM predictions to data.
  Figure~1(a) in the Letter is the combination of these two plots.
  Figure~\ref{fig:MorePlots} shows comparisons between data and
  simulation for other kinematic variables after the fit.
  
\end{document}

%% file: author_list.tex
\affiliation{Universidad de Buenos Aires, Buenos Aires, Argentina}
\affiliation{LAFEX, Centro Brasileiro de Pesquisas F{\'\i}sicas, Rio de Janeiro, Brazil}
\affiliation{Universidade do Estado do Rio de Janeiro, Rio de Janeiro, Brazil}
\affiliation{Universidade Federal do ABC, Santo Andr\'e, Brazil}
\affiliation{Instituto de F\'{\i}sica Te\'orica, Universidade Estadual Paulista, S\~ao Paulo, Brazil}
\affiliation{Simon Fraser University, Vancouver, British Columbia, and York University, Toronto, Ontario, Canada}
\affiliation{University of Science and Technology of China, Hefei, People's Republic of China}
\affiliation{Universidad de los Andes, Bogot\'{a}, Colombia}
\affiliation{Charles University, Faculty of Mathematics and Physics, Center for Particle Physics, Prague, Czech Republic}
\affiliation{Czech Technical University in Prague, Prague, Czech Republic}
\affiliation{Center for Particle Physics, Institute of Physics, Academy of Sciences of the Czech Republic, Prague, Czech Republic}
\affiliation{Universidad San Francisco de Quito, Quito, Ecuador}
\affiliation{LPC, Universit\'e Blaise Pascal, CNRS/IN2P3, Clermont, France}
\affiliation{LPSC, Universit\'e Joseph Fourier Grenoble 1, CNRS/IN2P3, Institut National Polytechnique de Grenoble, Grenoble, France}
\affiliation{CPPM, Aix-Marseille Universit\'e, CNRS/IN2P3, Marseille, France}
\affiliation{LAL, Universit\'e Paris-Sud, CNRS/IN2P3, Orsay, France}
\affiliation{LPNHE, Universit\'es Paris VI and VII, CNRS/IN2P3, Paris, France}
\affiliation{CEA, Irfu, SPP, Saclay, France}
\affiliation{IPHC, Universit\'e de Strasbourg, CNRS/IN2P3, Strasbourg, France}
\affiliation{IPNL, Universit\'e Lyon 1, CNRS/IN2P3, Villeurbanne, France and Universit\'e de Lyon, Lyon, France}
\affiliation{III. Physikalisches Institut A, RWTH Aachen University, Aachen, Germany}
\affiliation{Physikalisches Institut, Universit{\"a}t Freiburg, Freiburg, Germany}
\affiliation{II. Physikalisches Institut, Georg-August-Universit{\"a}t G\"ottingen, G\"ottingen, Germany}
\affiliation{Institut f{\"u}r Physik, Universit{\"a}t Mainz, Mainz, Germany}
\affiliation{Ludwig-Maximilians-Universit{\"a}t M{\"u}nchen, M{\"u}nchen, Germany}
\affiliation{Fachbereich Physik, Bergische Universit{\"a}t Wuppertal, Wuppertal, Germany}
\affiliation{Panjab University, Chandigarh, India}
\affiliation{Delhi University, Delhi, India}
\affiliation{Tata Institute of Fundamental Research, Mumbai, India}
\affiliation{University College Dublin, Dublin, Ireland}
\affiliation{Korea Detector Laboratory, Korea University, Seoul, Korea}
\affiliation{CINVESTAV, Mexico City, Mexico}
\affiliation{Nikhef, Science Park, Amsterdam, the Netherlands}
\affiliation{Radboud University Nijmegen, Nijmegen, the Netherlands and Nikhef, Science Park, Amsterdam, the Netherlands}
\affiliation{Joint Institute for Nuclear Research, Dubna, Russia}
\affiliation{Institute for Theoretical and Experimental Physics, Moscow, Russia}
\affiliation{Moscow State University, Moscow, Russia}
\affiliation{Institute for High Energy Physics, Protvino, Russia}
\affiliation{Petersburg Nuclear Physics Institute, St. Petersburg, Russia}
\affiliation{Instituci\'{o} Catalana de Recerca i Estudis Avan\c{c}ats (ICREA) and Institut de F\'{i}sica d'Altes Energies (IFAE), Barcelona, Spain}
\affiliation{Stockholm University, Stockholm and Uppsala University, Uppsala, Sweden}
\affiliation{Lancaster University, Lancaster LA1 4YB, United Kingdom}
\affiliation{Imperial College London, London SW7 2AZ, United Kingdom}
\affiliation{The University of Manchester, Manchester M13 9PL, United Kingdom}
\affiliation{University of Arizona, Tucson, Arizona 85721, USA}
\affiliation{University of California Riverside, Riverside, California 92521, USA}
\affiliation{Florida State University, Tallahassee, Florida 32306, USA}
\affiliation{Fermi National Accelerator Laboratory, Batavia, Illinois 60510, USA}
\affiliation{University of Illinois at Chicago, Chicago, Illinois 60607, USA}
\affiliation{Northern Illinois University, DeKalb, Illinois 60115, USA}
\affiliation{Northwestern University, Evanston, Illinois 60208, USA}
\affiliation{Indiana University, Bloomington, Indiana 47405, USA}
\affiliation{Purdue University Calumet, Hammond, Indiana 46323, USA}
\affiliation{University of Notre Dame, Notre Dame, Indiana 46556, USA}
\affiliation{Iowa State University, Ames, Iowa 50011, USA}
\affiliation{University of Kansas, Lawrence, Kansas 66045, USA}
\affiliation{Kansas State University, Manhattan, Kansas 66506, USA}
\affiliation{Louisiana Tech University, Ruston, Louisiana 71272, USA}
\affiliation{Boston University, Boston, Massachusetts 02215, USA}
\affiliation{Northeastern University, Boston, Massachusetts 02115, USA}
\affiliation{University of Michigan, Ann Arbor, Michigan 48109, USA}
\affiliation{Michigan State University, East Lansing, Michigan 48824, USA}
\affiliation{University of Mississippi, University, Mississippi 38677, USA}
\affiliation{University of Nebraska, Lincoln, Nebraska 68588, USA}
\affiliation{Rutgers University, Piscataway, New Jersey 08855, USA}
\affiliation{Princeton University, Princeton, New Jersey 08544, USA}
\affiliation{State University of New York, Buffalo, New York 14260, USA}
\affiliation{Columbia University, New York, New York 10027, USA}
\affiliation{University of Rochester, Rochester, New York 14627, USA}
\affiliation{State University of New York, Stony Brook, New York 11794, USA}
\affiliation{Brookhaven National Laboratory, Upton, New York 11973, USA}
\affiliation{Langston University, Langston, Oklahoma 73050, USA}
\affiliation{University of Oklahoma, Norman, Oklahoma 73019, USA}
\affiliation{Oklahoma State University, Stillwater, Oklahoma 74078, USA}
\affiliation{Brown University, Providence, Rhode Island 02912, USA}
\affiliation{University of Texas, Arlington, Texas 76019, USA}
\affiliation{Southern Methodist University, Dallas, Texas 75275, USA}
\affiliation{Rice University, Houston, Texas 77005, USA}
\affiliation{University of Virginia, Charlottesville, Virginia 22901, USA}
\affiliation{University of Washington, Seattle, Washington 98195, USA}
\author{V.M.~Abazov} \affiliation{Joint Institute for Nuclear Research, Dubna, Russia}
\author{B.~Abbott} \affiliation{University of Oklahoma, Norman, Oklahoma 73019, USA}
\author{B.S.~Acharya} \affiliation{Tata Institute of Fundamental Research, Mumbai, India}
\author{M.~Adams} \affiliation{University of Illinois at Chicago, Chicago, Illinois 60607, USA}
\author{T.~Adams} \affiliation{Florida State University, Tallahassee, Florida 32306, USA}
\author{G.D.~Alexeev} \affiliation{Joint Institute for Nuclear Research, Dubna, Russia}
\author{G.~Alkhazov} \affiliation{Petersburg Nuclear Physics Institute, St. Petersburg, Russia}
\author{A.~Alton$^{a}$} \affiliation{University of Michigan, Ann Arbor, Michigan 48109, USA}
\author{G.~Alverson} \affiliation{Northeastern University, Boston, Massachusetts 02115, USA}
\author{G.A.~Alves} \affiliation{LAFEX, Centro Brasileiro de Pesquisas F{\'\i}sicas, Rio de Janeiro, Brazil}
% ...
%
%
\collaboration{The D0 Collaboration\footnote{with visitors from
$^{a}$Augustana College, Sioux Falls, SD, USA,
$^{b}$The University of Liverpool, Liverpool, UK,
$^{c}$SLAC, Menlo Park, CA, USA,
$^{d}$University College London, London, UK,
$^{e}$Centro de Investigacion en Computacion - IPN, Mexico City, Mexico,
$^{f}$ECFM, Universidad Autonoma de Sinaloa, Culiac\'an, Mexico,
and 
$^{g}$Universit{\"a}t Bern, Bern, Switzerland.
$^{\ddag}$Deceased.
}} \noaffiliation
\vskip 0.25cm

%% file: acknowledgement.tex
%
We thank the staffs at Fermilab and collaborating institutions,
and acknowledge support from the
DOE and NSF (USA);
CEA and CNRS/IN2P3 (France);
FASI, Rosatom and RFBR (Russia);
CNPq, FAPERJ, FAPESP and FUNDUNESP (Brazil);
DAE and DST (India);
Colciencias (Colombia);
CONACyT (Mexico);
KRF and KOSEF (Korea);
CONICET and UBACyT (Argentina);
FOM (The Netherlands);
STFC and the Royal Society (United Kingdom);
MSMT and GACR (Czech Republic);
CRC Program and NSERC (Canada);
BMBF and DFG (Germany);
SFI (Ireland);
The Swedish Research Council (Sweden);
and
CAS and CNSF (China).

%% file: arxiv.bbl
\begin{thebibliography}{99}
\bibitem{bib:def}	D0 uses a spherical coordinate system with the $z$ axis running along the proton beam axis. The angles $\theta$ and $\phi$
are the polar and azimuthal angles, respectively. Pseudorapidity is defined as $\eta = -\ln \left[\tan(\theta/2) \right]$, in which $\theta$ is measured with respect to the proton beam direction.
\end{thebibliography}
